\documentclass[a4paper]{jpconf}
\usepackage{graphicx}
\newcommand{\nuc}[2]{\hbox{$^{#1}$#2}}
\begin{document}
\title{News on the nuclear structure of neutron-rich nuclei at and beyond N=28}

\author{Alexandra Gade}

\address{National Superconducting Cyclotron Laboratory and Department
  of Physics and Astronomy, Michigan State University, East Lansing,
  MI 48824, USA} 

\ead{gade@msu.edu}

\begin{abstract}
The nuclear potential and resulting shell 
structure are well established for the valley of stability, however, dramatic
modifications to the familiar ordering of single-particle orbitals
in rare isotopes with a large imbalance of proton and neutron numbers
have been found: new shell gaps emerge and conventional magic numbers
are no longer valid. This article outlines some of the recent in-beam $\gamma$-ray spectroscopy measurements at NSCL aimed at
shedding light on the evolution of nuclear structure around neutron number $N=28$ in
neutron-rich Ar and S isotopes. \\ \\
(Received April 2, 2018)
\end{abstract}

\section{Introduction}

Experiments on neutron-rich Si, S, and Ar isotopes continue to revolutionize our
understanding of the changes in nuclear structure encountered at the extremes of
isospin~\cite{Uts12,Now09,Sor13,Gad16a}, starting with the observation that the
$N=28$ shell closure may not persist for neutron-rich nuclei more than 20 years
ago~\cite{Sch96,Gla97} and continuing today with the discovery of
shape~\cite{For10} and configuration coexistence~\cite{San11,Uts15,Egi16,Par17}
in \nuc{44}{S}, for example.  

\begin{figure}[h]
\includegraphics[width=28pc]{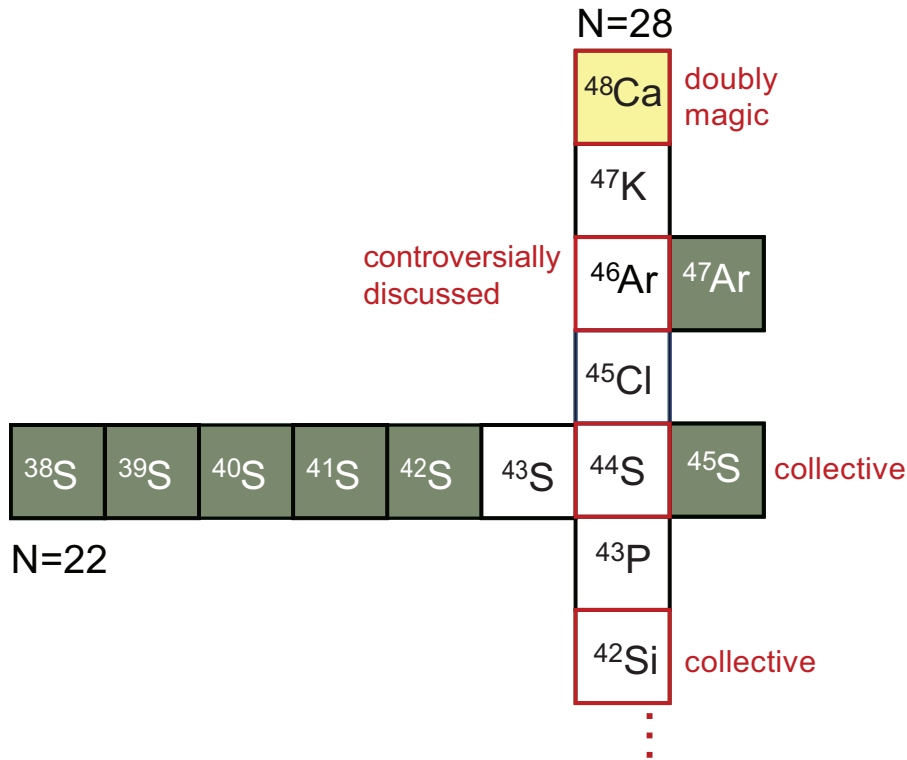}\hspace{2pc}%
\begin{minipage}[b]{\textwidth}\caption{(Color online) Rapid shell
    evolution has been 
    observed along the $N=28$ isotone line south of doubly magic
    \nuc{48}{Ca} as \nuc{42}{Si} is approached. Ar and S isotopes recently
    studied at NSCL are shaded and are the topic of this article.}   
\end{minipage}
\label{fig:intro}
\end{figure}

Recent experiments at the NSCL~\cite{Gad16b} used in-beam $\gamma$-ray
spectroscopy with fast beams of rare isotopes~\cite{Gad08} to explore, for key
isotopes or isotopic chains, the single-particle
structure from direct reactions or the detailed level structure as populated in
fragmentation reactions, respectively. Figure~\ref{fig:intro} highlights, on the
nuclear chart, the nuclei for which the results are reviewed in the present
article. 

One way of probing nuclear structure in a quantitative way is through the use of
direct nuclear reactions that selectively probe specific degrees of freedom,
e.g. the proton and neutron single-particle degrees of freedom. Intriguing
possibilities arise when the single-particle and single-hole structure of a
nucleus can be probed for protons and neutrons, using, for example, fast-beam
one-nucleon pickup~\cite{McD12,Gad11,Gad07} and one-nucleon-knockout
\cite{Han03} reactions with $\gamma$-ray detection to deduce the partial cross
sections for the population of individual final states. Other less selective
reactions, as for example secondary fragmentation of fast projectiles, 
populate the low-lying excited states of nuclei and can provide a first glimpse
at the excitation level schemes of exotic nuclei~\cite{Gad08}.

At NSCL, for the measurements described in the following, the $\gamma$-ray
spectroscopy was performed with either GRETINA~\cite{gretina} or
SeGA~\cite{sega} in coincidence with the detection of the projectile-like
reaction residues in the S800 spectrograph~\cite{s800}. The $\gamma$ rays
emitted by the projectiles in flight ($v/c \approx 0.3$) are detected with large
Doppler shifts in the laboratory frame and the segmentation or position
sensitivity of the $\gamma$-ray spectroscopy arrays is needed to event-by-event
Doppler reconstruct the $\gamma$-ray transitions into the reference frame of the
reaction residue. All reaction residues are identified and characterized using
the particle's time of flight and energy loss as measured with beam line timing
scintillators and the S800 focal-plane detection system.      

The following sections briefly summarize recent experiments performed at NSCL:
The study of \nuc{47}{Ar} in the one-neutron pickup reaction from \nuc{46}{Ar}
and the one-proton knockout from \nuc{48}{K}~\cite{Gad16c}, the first
observation of excited states in \nuc{45}{S}, and the detailed study of
excited-state level schemes of neutron-rich S isotopes approaching $N=28$ from
secondary fragmentation.   

\section{Learning from $N=29$ -- A complementary approach}
The role of the Ar isotopes around $N=28$ is of great
interest. They are, with element number $Z=18$, located between
doubly-magic \nuc{48}{Ca} and the already collective S isotopes~\cite{Gla97}
($Z=16$) on the path toward \nuc{42}{Si}~\cite{Bas07}, which has the
lowest-lying $2^+_1$ state along the $N=28$ isotonic chain. 

The nucleus \nuc{46}{Ar} has challenged existing shell model calculations which overpredict the
$B(E2)$ excitation strength to the first $2^+$
state~\cite{Sch96,Gad03,Win12,Cal16}, while describing well other 
observables, most recently related to masses of the Ar isotopes around $N=28$,
for example~\cite{Mei15}. This is not without 
experimental controversy, where an excited-state lifetime
measurement yields higher collectivity in  agreement with shell
model calculations~\cite{Men10}, but disagrees with several Coulomb 
excitation measurements that consistently yield a lower $B(E2)$
value~\cite{Sch96,Gad03,Cal16}. 

Reminded in the following is the study of \nuc{47}{Ar} using two complementary
intermediate-energy nuclear reactions,
\nuc{12}{C}(\nuc{46}{Ar},\nuc{47}{Ar}+$\gamma$)X ,  
and 
\nuc{9}{Be}(\nuc{48}{K},\nuc{47}{Ar}+$\gamma$)X and the first
spectroscopy of \nuc{45}{S} accomplished with a one-proton removal reaction from
\nuc{46}{Cl} projectiles~\cite{Gad16c}.    

\subsection{Single-neutron particle and single-proton hole strength}

We have presented a comprehensive spectroscopy of \nuc{47}{Ar} performed using two
complementary direct reactions, one-neutron pickup onto \nuc{46}{Ar} projectiles
and one-proton removal from the $1^-$ ground state of \nuc{48}{K}. Both of the
state-of-the-art SDPF-U and SDPF-MU shell-model effective interactions provided
a good description of the data, with (a) SDPF-U in better agreement with the
\nuc{47}{Ar} excitation energies and the location of the neutron spectroscopic
strength, and (b) SDPF-MU describing better the \nuc{47}{Ar} yield from proton
removal from the $2s_{1/2}$ orbital in the \nuc{48}{K} ground state~\cite{Gad16c}. Figure~2 summarizes the results. 

\begin{figure}[h]
\includegraphics[width=\textwidth]{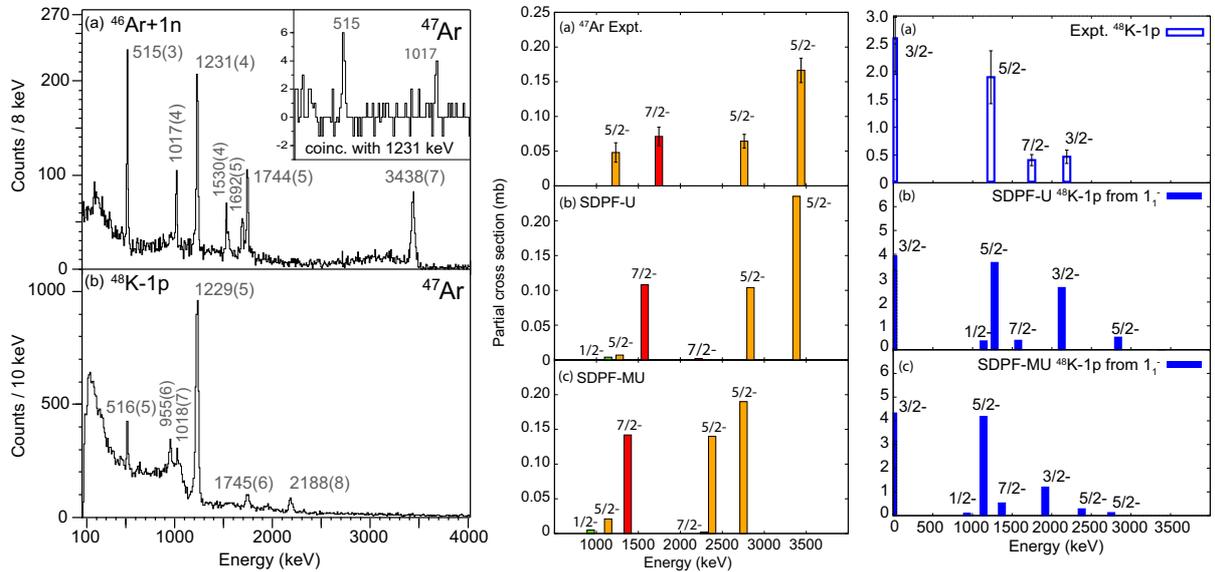}\hspace{2pc}%
\begin{minipage}[b]{\textwidth}\caption{(Color online) Left: \nuc{47}{Ar}
    $\gamma$-ray spectra as observed following the (a) one-neutron pickup onto
    \nuc{46}{Ar} projectiles (using GRETINA) and (b) one-proton removal from
    \nuc{48}{K} projectiles (using SeGA). Middle and right: comparison of
    measured and calculated cross sections using nuclear structure input from
    two different effective shell model interactions. Details are given in
    Ref.~\cite{Gad16c}. Figure adapted from~\cite{Gad16c}.}   
\end{minipage}
\label{fig:argon}
\end{figure}

\subsection{Reaching the furthest in the chain of S isotopes}
From the \nuc{9}{Be}(\nuc{46}{Cl},\nuc{45}{S}+$\gamma$)X one-proton removal
reaction, the first observation of $\gamma$-ray transitions in \nuc{45}{S} was
accomplished. This makes \nuc{45}{S} the most neutron-rich odd-$A$ sulfur isotope with spectroscopic information available, while \nuc{46}{S} is still the heaviest S isotope successfully accessed with $\gamma$-ray spectroscopy~\cite{Gad09} to date.
From
comparisons with shell model calculations, and arguments based on intensities
and energy sums, a first tentative level scheme for \nuc{45}{S} was proposed.
The \nuc{45}{S} $\gamma$-ray spectrum is broadly consistent with expectations
when assuming a $1^-$ \nuc{46}{Cl} ground-state spin and the the corresponding
proton spectroscopic factors from the SDPF-U interaction. Opportunities to advance our
understanding 
of the $N=29$ nucleus \nuc{45}{S} will emerge once the ground-state spin of
\nuc{46}{Cl} is known, after which a more quantitative analysis, as done for
\nuc{47}{Ar}, can be performed for the one-proton removal data~\cite{Gad16c}. Figure~3 shows the $\gamma$-ray spectrum as Doppler-reconstructed with SeGA and the particle identification spectrum that used information from the S800 spectrograph.

\begin{figure}[h]
\includegraphics[width=\textwidth]{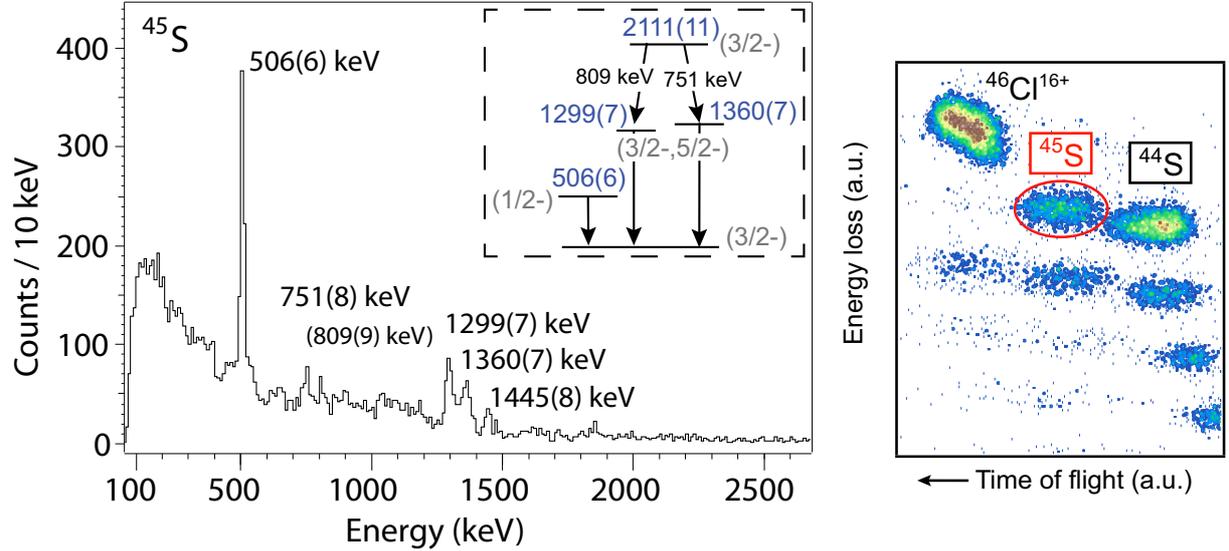}\hspace{2pc}%
\begin{minipage}[b]{\textwidth}\caption{(Color online) Left: Gamma-ray spectrum observed with
    SeGA in coincidence with \nuc{45}{S} and tentative level scheme. No $\gamma$-ray transitions were known
    before. Right: Particle identification spectrum of \nuc{45}{S}. 
    Details are given in Ref.~\cite{Gad16c}. Figure adapted from~\cite{Gad16c}.}    
\end{minipage}
\label{fig:sulfur45}
\end{figure}

\section{Approaching $N=28$ in the chain of S isotopes}

We have performed in-beam $\gamma$-ray spectroscopy on neutron-rich sulfur isotopes
  populated by fragmentation of intermediate-energy \nuc{48}{Ca} and
  \nuc{46}{Ar} projectile beams. New transitions were identified in
  \nuc{39-42}{S} and new level schemes for \nuc{40-42}{S} were proposed from
  $\gamma\gamma$ coincidence information, energy sums and comparison to shell
  model. Shell-model calculations with the SDPF-MU Hamiltonian provideed
  remarkable agreement and consistency with the proposed level schemes. For the
odd-mass \nuc{41}{S}, a level scheme is presented that appears complete below
2.2~MeV and consistent with the predictions by SDPF-MU
shell-model Hamiltonian (see Figure~4); this is a remarkable benchmark given the rapid shell
and shape evolution prevalent in this textbook isotopic chain as the diminished
$N=28$ shell gap is approached. 

\begin{figure}[h]
\includegraphics[width=\textwidth]{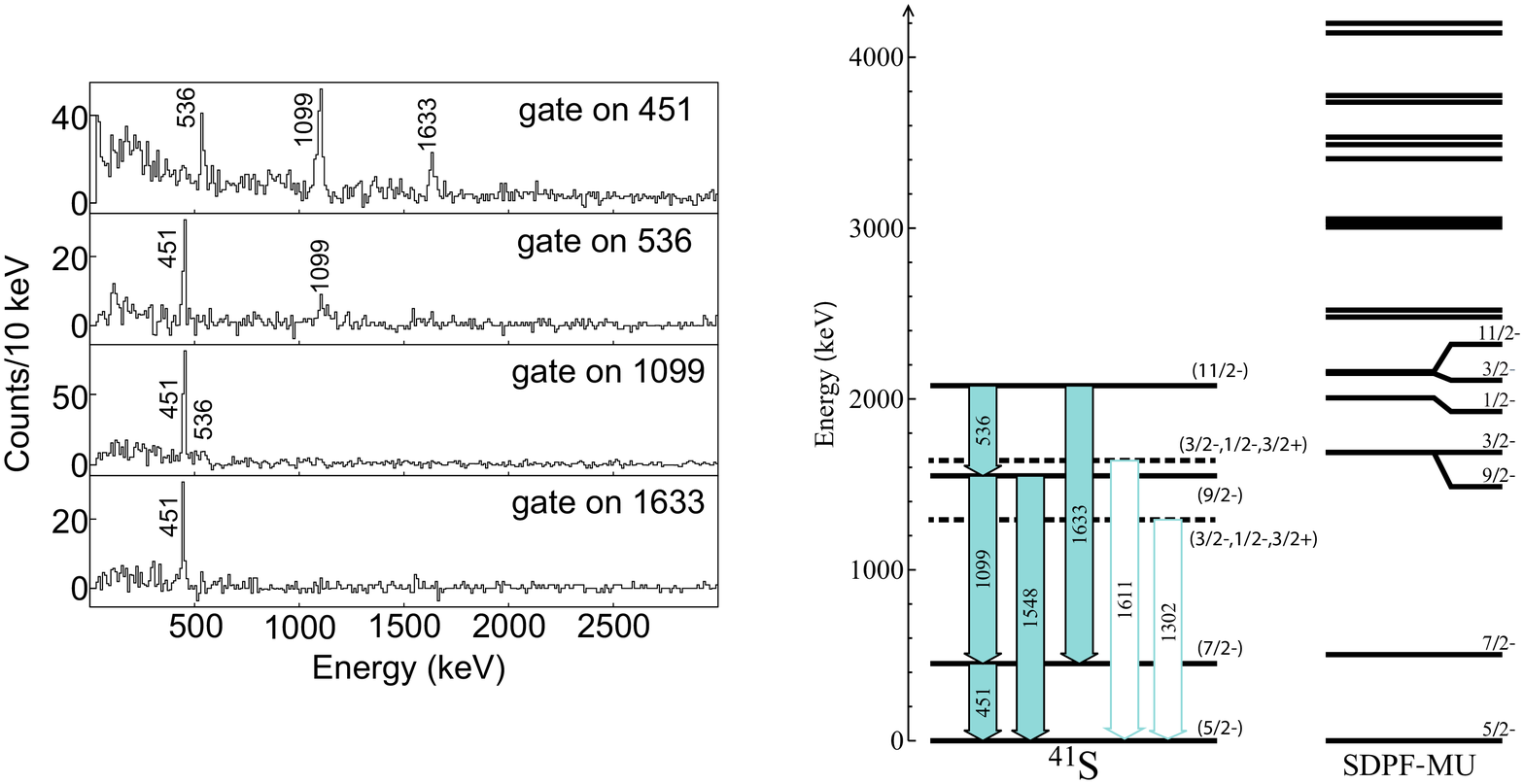}\hspace{2pc}%
\begin{minipage}[b]{\textwidth}\caption{(Color online) Gamma-ray coincidence spectra for
    \nuc{41}{S} (left) and resulting level scheme compared to shell model
    calculations (right). Unfilled arrow indicate tentative placement of transitions. Details are given in Ref.~\cite{Lun16}. Figure adapted
    from~\cite{Lun16}.}     
\end{minipage}
\label{fig:sulfurs}
\end{figure}

As detailed in Ref.~\cite{Lun16}, for the even-mass S isotopes, the evolution of
the yrast sequence was discussed in terms of $E(6^+)/E(2^+)$ and $E(4^+)/E(2^+)$
energy ratios. For \nuc{42}{S}, a candidate for the $2^+_2$ state could be
proposed that displays a unique decay branching as compared to the lighter
\nuc{38,40}{S}. This was shown to be rooted in its neutron single-particle
structure and confirmed by the SDPF-MU shell-model calculations.  

As to the population of excited states in fragmentation reactions, a consistent
picture emerged. Transitions from yrast states were most prominent, evident even
at low statistics. For the higher statistics cases, e.g. \nuc{40,41,42}{S}, the
presence of many weaker transitions became apparent that most certainly feed
the low-lying level schemes. While this may always have been the assumption
behind the population of excited states in fragmentation reactions, evidence was
presented for the many feeding transitions that have remained unobserved in
previous work discussing fragmentation reactions specifically for S
isotopes~\cite{Soh02}. In the case of \nuc{41}{S}(\nuc{42}{S}), all calculated
negative(positive) parity states below 2.2~MeV(3.5~MeV) were matched to states
of the proposed level schemes, including off-yrast states. A number of weaker
transitions remained unplaced.

\section{Summary}
Neutron-rich Ar and S nuclei continue to provide important benchmarks for nuclear theory in the quest to unravel the driving forces of shell and shape evolution. The results from recent in-beam $\gamma$-ray spectroscopy studies on \nuc{47}{Ar}, \nuc{45}{S}, and \nuc{38-42}{S} were reviewed. In general, very good agreement between measured results and large-scale shell model calculations  was demonstrated.

\section*{Acknowledgments}
This work was supported by the National Science Foundation under
Grants No. PHY-1565546. Over the many years, more than a decade by now, valuable collaboration and countless
inspiring discussions with Takaharu Otsuka are acknowledged that have shaped the
in-beam $\gamma$-ray spectroscopy program at NSCL.

\end{document}